\RequirePackage{fix-cm}
\documentclass[smallextended]{svjour3}       
\smartqed  
\usepackage{tablefootnote}
\usepackage{authblk}
\usepackage{amsmath, amsfonts, amssymb, color, graphicx, mathtools, enumerate, subfigure}
\usepackage{color}
\usepackage{bm}
\usepackage{algorithmicx}
\usepackage{algpseudocode}
\usepackage{booktabs,caption}
\usepackage[flushleft]{threeparttable}
\usepackage{setspace}
\usepackage{caption} 
\usepackage[utf8]{inputenc}
\usepackage[linesnumbered,algoruled,lined]{algorithm2e}

\usepackage{multirow, float, changepage}
\usepackage{natbib}
\setcitestyle{authoryear,round}
\usepackage{wrapfig}
\usepackage{changes}

\usepackage[CJKbookmarks=true,
            bookmarksnumbered=true,
  bookmarksopen=true,
  colorlinks=true,
  citecolor=blue,
  linkcolor=blue,
  anchorcolor=blue,
  urlcolor=blue]{hyperref}
  
\usepackage[letterpaper, left=1.2truein, right=1.2truein, top = 1.2truein, bottom = 1.2truein]{geometry}
\usepackage{lipsum}
\usepackage{filecontents}

\usepackage{caption}

\providecommand{\keywords}[1]
{
  {\small	
  \textbf{{Keywords:}} #1
  }
}
\usepackage{prettyref,soul}
\usepackage{varwidth}

\begin{document}

\title{Randomized Kaczmarz Method for Single Particle X-ray Image Phase Retrieval
}


\author{Yin Xian         \and
        Haiguang Liu \and
        Xuecheng Tai \and
        Yang Wang 
}


\institute{Yin Xian \at
              Hong Kong University of Science and Technology, Hong Kong SAR China. \\
              \email{poline3939@gmail.com}  
		\and
		Haiguang Liu \at
              Microsoft Research Lab - Asia, Beijing, China. \\
              \email{haiguangliu@microsoft.com}           
           \and
           Xuecheng Tai \at
           Hong Kong Baptist University, Hong Kong SAR China. \\
           \email{xuechengtai@hkbu.edu.hk}           
           \and
           Yang Wang \at
           Hong Kong University of Science and Technology, Hong Kong SAR China. \\
           \email{yangwang@ust.hk}          
}

\date{Received: date / Accepted: date}

\maketitle

\tableofcontents

\vspace{15pt}


\begin{abstract}
 In this chapter, we investigate phase retrieval algorithm for the single particle X-ray imaging data. We present a variance-reduced randomized Kaczmarz (VR-RK) algorithm for phase retrieval. The VR-RK algorithm is inspired by the randomized Kaczmarz method and the Stochastic Variance Reduce Gradient Descent (SVRG) algorithm. Numerical experiments show that the VR-RK algorithm has a faster convergence rate than randomized Kaczmarz algorithm and the iterative projection phase retrieval methods, such as the hybrid input output (HIO) and the relaxed averaged alternating reflections (RAAR) methods. The VR-RK algorithm can recover the phases with higher accuracy, and is robust at the presence of noise. Experimental results on the scattering data from individual particles show that the VR-RK algorithm can recover phases and improve the single particle image identification. 
 


\end{abstract}

\section{\textbf{Introduction}}
\label{intro}
\subsection{\textbf{The Phase Retrieval Problem}}

The mathematical formulation of phase retrieval is solving a set of quadratic equations. Methods to solve the phase retrieval problem can be classified into two categories: convex and non-convex approaches. Convex methods like \textit{PhaseLift}~(\cite{candes2013}), convert the quadratic system equation to a linear system equation through a matrix-lifting technique. The \textit{PhaseMax} method~(\cite{goldstein2017, bahmani2017}) operates in the original signal space rather than lifting it to a higher dimensional space. It replaces the non-convex constraints with inequality constraints that define convex sets. The convex approaches have good recovery guarantees, but their computational complexities are usually high when the dimension of the signals is large. 

On the other hand, the non-convex approaches turn the phase retrieval into an optimization problem. The most popular class of methods is based on alternate projection, such as the hybrid input output (HIO) method~(\cite{bauschke2003}), the Error Reduction (ER) method~(\cite{fienup1986}) and the relaxed averaged alternating reflections (RAAR) method~(\cite{luke2005}). These methods are iterative projection methods, since they involve iterative projections onto the constraint sets.  Unlike the convex approaches, convergence is not guaranteed for these algorithms, and stagnation may occur due to non-uniqueness of the solution~(\cite{fienup1986}). A unified evaluation of these iterative projection algorithms can be found in the paper of~\cite{marchesini2007}. Recently, a method called Wirtinger flow~(\cite{candes2015}) is proposed. It works well with spectral method for initialization. The follow-up works include the truncated Wirtinger flow~(\cite{chen2015}), truncated amplitude flow~(\cite{wang2017}) and reshaped Wirtinger flow~(\cite{zhang2016}).  These methods have less computational complexities and have theoretical convergence guarantees.


The randomized Kaczmarz algorithm is introduced to solve the phase retrieval problem by Wei~(\cite{wei2015}).The randomized Kaczmarz method can be viewed as a special case of the stochastic gradient descent (SGD)~(\cite{needell2014}). For the phase retrieval problem, the method is essentially SGD for the amplitude flow objective. It was shown numerically that the method outperforms the Wirtinger flow and the ER method~(\cite{wei2015}). The convergence rate of the randomized Kaczmarz method for the linear system is studied in the paper of \cite{strohmer2009}. The theoretical justification of using randomized Kaczmarz method for phase retrieval has been presented in the paper of \cite{tan2018}. 


\subsection{\textbf{Challenges of X-ray Data Processing}}

The structure of biological macromolecules is the key to understand the living cells function and behavior. The Protein Data Bank (PDB)~(\cite{bernstein1977}) currently has more than 173,110 structures, but many structures of biological molecules and their complexes have not been determined. The cryo-electron microscopy (Cryo-EM) and the X-ray crystallography have been successfully applied in this field. The X-ray crystallography has solved about 90\% of these structures. However, growing high quality crystals of biomolecules is challenging, especially for biologically functional molecules. Therefore, determining structures from single molecules are appealing.

The use of the X-ray free electron lasers (XFEL) is a recent development in structure biology. The idea behind this method is to record the instantaneous elastic scattering from an ultrashort pulse. The pulse is so brief that it terminates before the onset of radiation damage (``diffract before destroy")~(\cite{liu2016}).  With this application, the single particle imaging becomes possible, even at room temperature. It allows one to understand the structures and dynamics of macromolecules.  

The difference between the Cryo-EM and the X-ray crystallography is that the Cryo-EM data includes phase information of the structural factors, while the X-ray crystallographic diffraction data only provide amplitude information, but lack phase information~(\cite{wangh2017, scheres2012}). The illustrations and data processing examples are shown in the paper of~\cite{sorzano2004, xian2018}, and~\cite{gu2020gen}. In order to solve the biological structures, the phase information is essential. It is normally obtained by experimental or computational means. 

The challenges of XFEL single particle imaging also include the following: (i) the signal-to-noise ratio (SNR) is low, and the information is  influenced by noise; (ii) the orientation of each sample particle is unknown, leading to the difficulty in data merging and 3D reconstruction; (iii) conformational heterogeneity places a hurdle for single particle identification and reconstruction~(\cite{wangh2017}).  In this chapter, we investigate the phase retrieval algorithms of the XFEL data. The baseline for a good phase retrieval algorithm is its robustness against noises and the incompleteness of information ~(\cite{shi2019}). 

\subsection{\textbf{Phase retrieval with noisy or incomplete measurements}}
The number of photons detected by the optical sensor is of Poisson distribution. For the phase retrieval problem contaminated by the Poisson noise, or has incomplete magnitude information, the prior information is crucial to process the data. Research for imposing prior information to image processing is shown in the literature ~(\cite{le2007variational,zhang2012novel,hunt2018data}).

In order to better reconstruct the data, one can consider a variational model by introducing a total variation (TV) regularization, which is widely used in imaging processing community. TV regularization can enable recovery of signals from incomplete or limited measurements. The alternating direction of multipliers method (ADMM)~(\cite{glowinski1989augmented, wu2010augmented}), and the split Bregman method~(\cite{goldstein2009split}) are usually applied to solve the TV-regularization problem. They have been applied in the phase retrieval problem~(\cite{chang2016phase, chang2018total, bostan2014phase, li2016total}).
 
Besides TV regularization, Tikhonov regularization is another important smoothing techniques in variational image denoising. It is often applied in noise removal. The phase retrieval problem with a Tikhonov regularization has been solved by the Gauss-Newton method~(\cite{seifert2006multilevel, sixou2013absorption, langemann2008phase, ramos2019direct}). Considering the sparsity constraints, the fixed point iterative approach~(\cite{fornasier2008recovery, tropp2006algorithms, ma2018implicit}) has been applied for the problem with nonlinear joint sparsity regulation.

\subsection{\textbf{Outline}}
In this chapter, we further advance the convergence speed of the randomized Kaczmarz method for phase retrieval. The idea comes from the fact that the randomized Kaczmarz method is a weighted SGD, and the convergence rate of SGD is slower because of the random sampling variance. Therefore, reducing the sampling variance can improve the convergence rate of the randomized Kaczmarz method. Inspired by the stochastic variance reduce gradient (SVRG) method~(\cite{johnson2013}), we present the variance-reduced randomized Kaczmarz method (VR-RK) for single particle X-ray imaging phase retrieval. Considering the sparsity constraint and generality of the problem, we present the VR-RK method under both the $L_1$ and the $L_2$ constraints for computational analysis. Numerical results on the virus data show that the VR-RK method can recover information with higher accuracy at a faster convergence rate. It helps recover the lost information due to the beam stop for blocking the incidence X-ray beam. 

The rest of the chapter is organized as follows. In section ``Background: Phase Retrieval and Stochastic Optimization", we give a general overview of phase retrieval and stochastic optimization. In section ``Variance-Reduced Randomized Kaczmarz (VR-RK) method", the proposed variance reduced randomized Kaczmarz method, and its variation under $L_1$ and $L_2$ constraints are presented. The evaluation of the algorithm is shown in the ``Application: Robust Phase Retrieval of the Single Particle X-ray Images" section, and the single particle X-ray image data are tested. The ``Conclusion" section concludes the chapter.
 
\section{\textbf{Background: Phase Retrieval and Stochastic Optimization}} \label{sec:background}

\subsection{\textbf{Phase Retrieval}}
\label{sec:phase_retrieval}
Formulation of the phase retrieval problem is as follows:
\begin{align}
\min\limits_{\bm{x}}\sum\limits_{k=1}^{m}(y_k-|\langle \bm{a_k,x} \rangle|^2)^2 .
\label{eq:phase_re_1}
\end{align}
where $\bm{y}$ is the measurement, $\bm{x}$ is the signal that need to be recovered, and $\bm{a_k}$ is the measurement operating vector. In the setting of forward X-ray scattering imaging at the far field, $\bm{a_k}$ is a Fourier vector, and $\bm{y}$ is a diffraction pattern of the target. The problem in phase retrieval is the limitation of optical sensors, which measures only the intensity. 

The loss function of eq.~(\ref{eq:phase_re_1}) is expressed as the squared difference between measurement intensities and the modelled intensities.  It is a system of quadratic equation, and therefore it is a non-convex problem.  

To solve eq.~(\ref{eq:phase_re_1}), the alternate projection methods are often used, such as HIO, ER, and RAAR methods as mentioned previously. These algorithms can be expressed in the form of fixed-point equation. They can be implemented jointly to better avoid local minima.  


When the loss function is expressed as the squared loss of amplitudes, the formulation can be written as:
\begin{align}
\min\limits_{\bm{x}}\sum\limits_{k=1}^{m}(\sqrt{y_k}-|\langle \bm{a_k,x} \rangle|)^2 .
\label{eq:phase_re_3}
\end{align}
To solve eq~(\ref{eq:phase_re_3}), it is possible to apply the amplitude flow algorithm~(\cite{wang2017}), which is essentially a gradient descent algorithm that can converge under good initialization.

\subsection{\textbf{Stochastic Optimization and the Kaczmarz Method}}
\label{sec:sgd}
The phase retrieval problem can be solved by stochastic optimization approaches. For the problem:
\begin{align}
\min\limits_{\bm{x}}\frac{1}{m}\sum\limits_{k=1}^{m}f_k(\bm{x}),
\end{align}
the gradient descent method updating rule is: $\bm{x_{k+1}}=\bm{x_k}-\frac{t_k}{m}\sum\limits_{k=1}^{m}\nabla f_k(\bm{x_k})$, where $t_k$ is the step size at each iteration, and $m$ is the number of samples, or the number of measurements in the phase retrieval setting. The gradient descent is expensive and it requires evaluation of $n$ derivatives at each iteration. To reduce the computational cost, the SGD is proposed:
\begin{align}
\bm{x_{k+1}}=\bm{x_k}-t_k\nabla f_{i_k}(\bm{x_k})
\label{eq:sgd}
\end{align}
where $i_{k}$ is an index chosen uniformly in random from $\{1,\cdots,m \}$ at each iteration. The computational cost is $1/m$ of the standard gradient descent. The SVRG is proposed to reduce variance of SGD, and has a faster convergence rate~(\cite{johnson2013}). It is operated in epochs. In each epoch, the updating process is:
\begin{align}
\bm{x_{k+1}}=\bm{x_k}-\eta \Bigl(\nabla f_{i_k}(\bm{x_k})-\nabla f_{i_k}(\bm{\bar{x}})+\frac{1}{m}\sum\limits_{i=1}^{m}\nabla f_i(\bm{\bar{x}}) \Bigr)
\label{eq:svrg}
\end{align}
where $\eta$ is the step size, and $\bm{\bar{x}}$ is a snapshot value in each epoch~(\cite{johnson2013}). 

The Kaczmarz method is a well-known iterative method for solving a system of linear equations $\bm{Ax=b}$, where $\bm{A}\in \mathbb{R}^{m\times n}$, $\bm{x}\in\mathbb{R}^n$, and $\bm{b}\in\mathbb{R}^m$. The classical Kaczmarz method sweeps through the rows in $\bm{A}$ in a cyclic manner, and projects the current estimate onto a hyperplane associated with the row of $\bm{A}$ to get the new estimate.  The randomized Kaczmarz method randomly chooses the row for projection in each iteration:
\begin{align} 
\bm{x_{k+1}}=\bm{x_k}+\frac{b_{i_k}-\langle \bm{a_{i_k}}, \bm{x_k} \rangle}{||\bm{a_{i_k}}||_2^2}\bm{a_{i_k}}
\label{eq:kaczmarz}
\end{align}
where $\bm{a_{i_k}}$ is the row of $\bm{A}$. The randomized Kaczmarz can be viewed as a reweighted SGD with importance sampling for the least squares problem~(\cite{needell2014}):
\begin{align}
F(\bm{x})=\frac{1}{2}||\bm{Ax-b}||_2^2=\frac{1}{2}\sum\limits_{i=1}^{m}(\bm{a_i^T}\bm{x}-b_i)^2.
\label{eq:least_sqrt}
\end{align}

The randomized Kaczmarz algorithm is essentially stochastic gradient descent for the amplitude flow problem in eq.~(\ref{eq:phase_re_3}). This suggests that the acceleration schemes for SGD, such as the variance reduced approach, can be applied to the algorithm, and improve phase retrieval.

\section{\textbf{Variance-Reduced Randomized Kaczmarz (VR-RK) method}} \label{sec:VR-RK}

Define $b_{i_k}=\sqrt{y_{i_k}}$, the formulation of eq.~(\ref{eq:phase_re_3}) can be written:
\begin{align}
\min\limits_{\bm{x}}\sum\limits_{k=1}^{m}(b_k-|\langle \bm{a_k,x} \rangle|)^2 .
\label{eq:phase_re_4}
\end{align}

The update scheme for randomized Kaczmarz for the phase retrieval objective of eq.~(\ref{eq:phase_re_4}), according to the paper of~\cite{tan2018}, is:
\begin{align}
\bm{x_{k+1}}=\bm{x_k}+\eta_k \bm{a_{i_k}} 
\label{eq:phase_rk_up}
\end{align}
where,
\begin{align*}
\eta_k=\frac{{\rm sign} (\langle \bm{a_{i_k}}, \bm{x_{k}} \rangle)b_{i_k} - \langle \bm{a_{i_k}}, \bm{x_{k}} \rangle}{||\bm{a_{i_k}}||_2^2} 
\end{align*} 
$i_k$ is drawn independently and identically distributed (i.i.d.) from the index set $\{1,2, \cdots, m \}$ with the probability  
\begin{align}
g_k=\frac{||\bm{a_{i_k}}||^2}{||\bm{A}||_F^2}.
\label{eq:prob}
\end{align}

The VR-RK method is inspired by the randomized Kaczmarz method and the SVRG method. It is proposed originally to solve the linear system equation~(\cite{jiao2017}). Let $f_i(\bm{x})=\frac{1}{2}(\bm{a_i^Tx}-b_i)^2$, and let
\begin{align}
h_i(\bm{x})=\frac{ f_i(\bm{x}) }{g_i}=\frac{1}{2}(\bm{|a_i^T x|}-b_i)^2\frac{||\bm{A}||_F^2}{||\bm{a_i}||^2}
\end{align}
then,
\begin{align}
\nabla h_i(\bm{x})=(\bm{a_i^Tx}-{\rm sign}(\bm{a_i^T x}) b_i)\bm{a_i} \frac{||\bm{A}||_F^2}{||\bm{a_i}||^2}
\end{align}
Let $\mu_i(\bm{x})=\nabla h_i(\bm{x})$, and $s$ be the size of the epoch. The Variance-Reduced Randomized Kaczmarz algorithm for phase retrieval is shown in Algorithm~1.
\begin{algorithm}[h]
\label{al:vr_rk}
\caption{Variance-Reduced Randomized Kaczmarz (VR-RK)}
Initialize $\mu_i(\bm{\bar{x}})=0$, and $\bar{\mu}=0$, specify $\bm{A}, \bm{b}, s$ \\
At steps $k=1,2,\cdots$,  if $k\mod s =0$, then
\begin{align*}
\bm{\bar{x}}=\bm{x_k} ~~\text{and  }~\bm{\bar{\mu}}=\mu(\bm{x_k})
\end{align*}
Pick index $i$ uniformly at random according to (\ref{eq:prob}). \\
Update $\bm{x_k}$ by
\resizebox{.9\linewidth}{!}{
 \begin{minipage}{\linewidth}
\begin{align*}
\bm{x_{k+1}}=\bm{x_k}-\frac{m}{||\bm{A}||_F^2}\bigl(\mu_{i_k}(\bm{x_k})-\mu_i (\bm{\bar{x}}) + \bm{\bar{\mu}}\bigr) 
\end{align*}
where $\bm{\bar{\mu}}=\frac{1}{m}\sum\limits_{i=1}^{m}\nabla h_i(\bm{\bar{x}})$
\end{minipage} }
\end{algorithm}

Considering the generality of the problem, and $L_2$ constraint is imposed, the objective function is:
\begin{align}
\min\limits_{\bm{x}}\frac{1}{2}\sum\limits_{k=1}^{m}(b_k-|\langle \bm{a_k,x} \rangle|)^2 +\gamma ||\bm{x}||_2.
\label{eq:vr_rk_l2}
\end{align}
Applying the randomized Kaczmarz method, according to~\cite{hefny2017}, the updating process becomes:
\begin{align}
\bm{x_{k+1}}=\bm{x_k}-\frac{(\bm{a_{i_k}}^T\bm{x_k}-{\rm sign}(\bm{a_{i_k}}^T\bm{x_k})b_{i_k})\bm{a_{i_k}}+\gamma \bm{x_k}}{||\bm{a_{i_k}}||^2+\gamma}
\end{align}

In the VR-RK setting, the updating process is 
\begin{align}
\nabla c_{i_k}(\bm{x_k})=\frac{(\bm{a_{i_k}}^T\bm{x_k}-{\rm sign}(\bm{a_{i_k}}^T\bm{x_k})b_{i_k})\bm{a_{i_k}}+\gamma \bm{x_k}}{||\bm{a_{i_k}}||^2+\gamma}    \\
\bm{x_{k+1}}=\bm{x_k}-\nabla c_{i_k}(\bm{x_k})+\nabla c_{i_k}(\bm{\bar{x}})-\frac{1}{m}\sum\limits_{i=1}^m \nabla c_i(\bm{\bar{x}})
\end{align}

For the consideration of the sparsity, the $L_1$ instead of the $L_2$ constraint can be imposed, then the objective function becomes:
\begin{align}
\min\limits_{\bm{x}}\frac{1}{2}\sum\limits_{k=1}^{m}(b_k-|\langle \bm{a_k,x} \rangle|)^2 +\lambda ||\bm{x}||_1.
\label{eq:vr_rk_l1}
\end{align}

To deal with this formula, the majorization-minimization (MM) technique and the C-PRIME method~(\cite{qiu2017}) is employed. It is shown that the problem is equivalent to:
\begin{align}
\min\limits_{\bm{x}}\biggl(C||\bm{x-d}||_2^2+\lambda||\bm{x}||_1  \biggr)
\label{eq:vr_rk_l1_2}
\end{align} 
where $C$ is a constant and $C\geq \rho_{max}(\bm{A}^H\bm{A}) $, $\rho_{max}$ is the largest eigenvalue of a matrix, and $\bm{d}$ is the constant vector that defined as:
\begin{align}
    \bm{d}&: =\bm{x_k}-\frac{1}{C}\bm{A}^H(\bm{A}\bm{x_k}-\bm{b}\odot e^{j\angle (\bm{A}\bm{x_k})}). 
\end{align}
Above, the notation $\odot$ is the element-wise Hadamard product of two vectors, and $\angle$ is the phase angle. The close form solution of $\bm{x}$ is:
\begin{align*}
    \bm{x}^{*}&=e^{j\angle (\bm{d})} \odot \max\left\{|\bm{d}|-\frac{\lambda}{2C}\bm{1},\bm{0}\right\} .
\end{align*}

\section{\textbf{Application: Robust Phase Retrieval of the Single Particle X-ray Images}} \label{sec:application}

In this section, we present numerical results of phase retrieval of the single particle X-ray imaging data. 

\subsection{\textbf{Synthetic single particle data recovery experiment}}
 \label{sec:syn_recovery}
\begin{figure}[ht]
    \centering
    \subfigure[]{\includegraphics[width=0.3\textwidth, height=36mm]{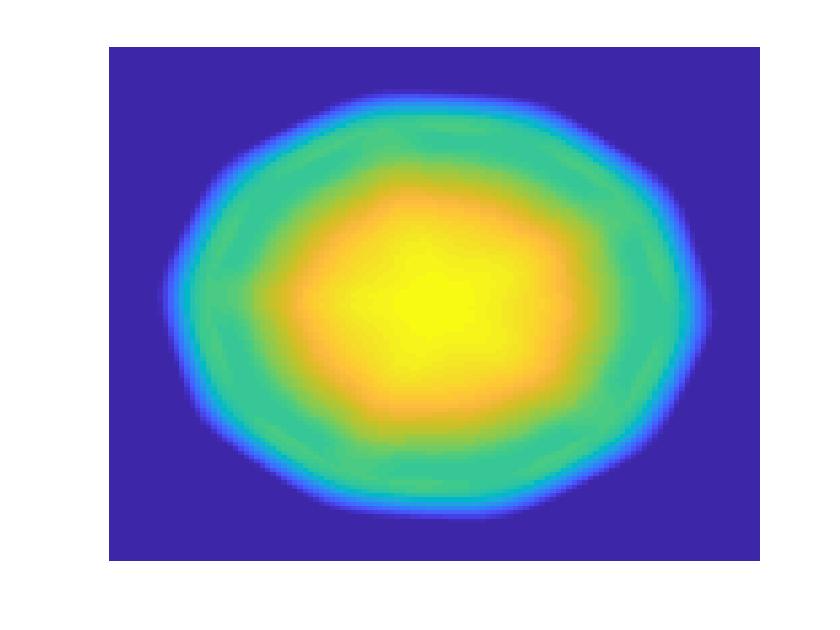}} 
    \subfigure[]{\includegraphics[width=0.3\textwidth, height=36mm]{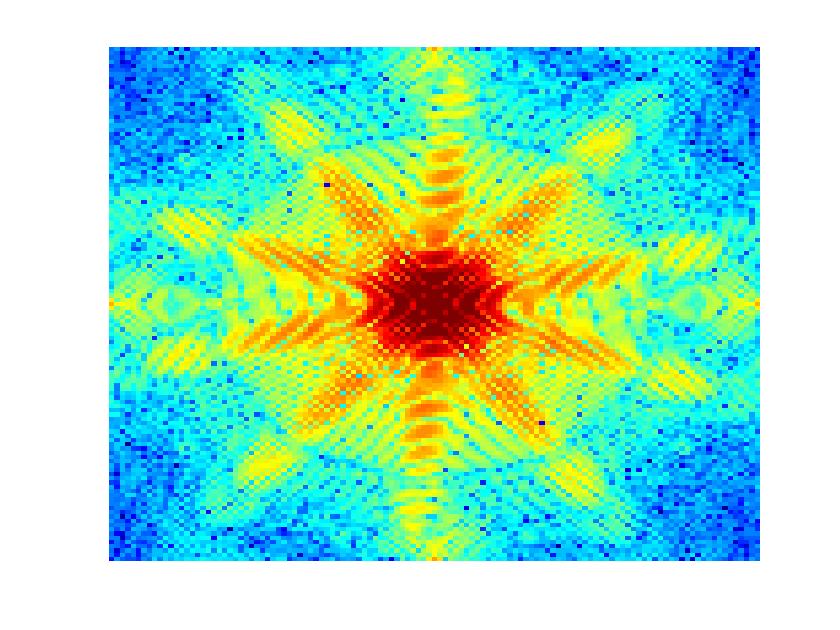}} 
      \subfigure[]{\includegraphics[width=0.3\textwidth, height=36mm]{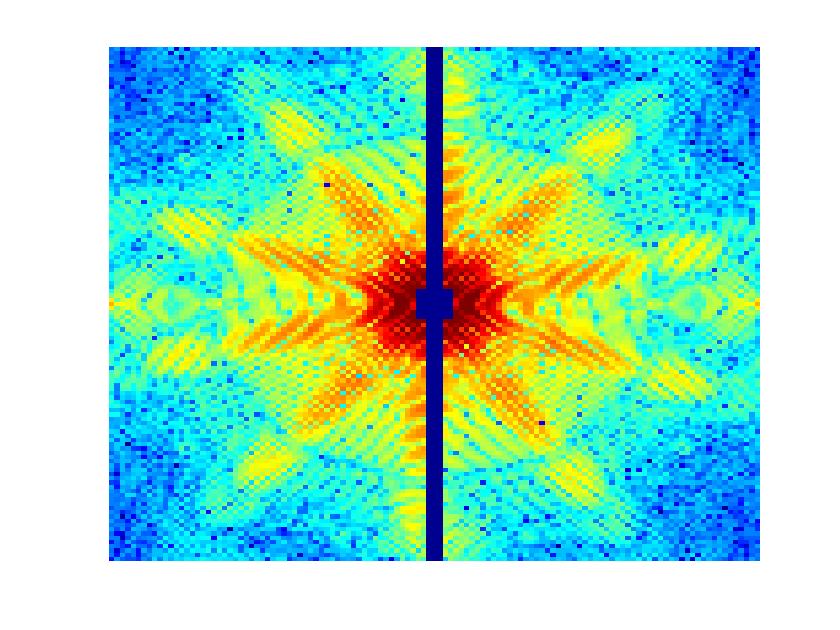}} 
\caption{Virus sample particle and its diffraction patterns~(\cite{li2016}). (a) Virus particle 2D projection imaging in real space. (b) Simulated X-ray data. (c) The simulated data with a gap. The size of pixels in the gap is 409. } 
    \label{fig:mimi_virus}
\vspace{-3mm}
\end{figure}

The first experiment is to test the reconstruction efficiency of the virus data, as shown in Figure~\ref{fig:mimi_virus}. The image size of Figure~\ref{fig:mimi_virus}(a) is 755 $\times$ 755 pixels, and the pixel values are normalized to [0,1]. The diffraction pattern (Figure~\ref{fig:mimi_virus}(b)) is created by taking the Fourier transform of Figure~\ref{fig:mimi_virus}(a). In this experiment, X-ray scattering signals are mainly observed at low resolutions, corresponding to low frequencies in Fourier space. A gap is placed in the center of the diffraction pattern to allow the incident beam to pass through, to avoid damaging or saturating detector sensors. The gap results in an information loss at low frequency regime, as shown in Figure~\ref{fig:mimi_virus}(c). The low frequency information corresponds to the overall shape of the object. Without which, it poses a challenge for reconstruction.    
\vspace{-0.15in}

\begin{figure}[ht]
    \centering
    \includegraphics[width=0.58\textwidth, height=48mm]{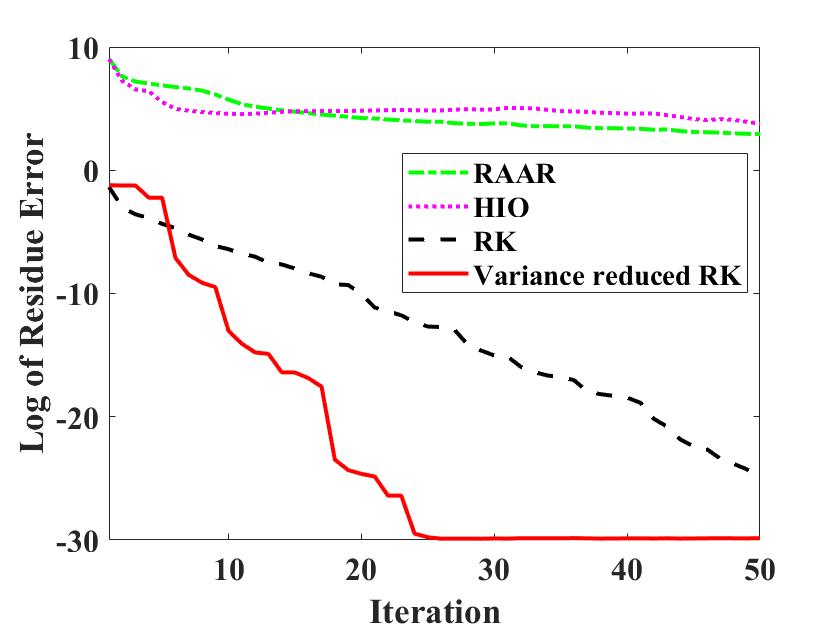}
\caption{Comparison of convergence rate}
    \label{fig:mimi_virus_converg}
\end{figure}

We reconstruct the sample virus image from the diffraction pattern with detector gap in Figure 1(c). The VR-RK, Randomized Kaczmarz, HIO, and RAAR methods are tested in the MATLAB platform. In order to reconstruct the data, a reference signal is used as  a \textit{priori} for preprocessing as described in the paper of~\cite{barmherzig2019}, and the numerical iteration is then performed. Comparison of convergence rate and the relative square errors are shown in Figure~\ref{fig:mimi_virus_converg} and  Table~\ref{table:mimi_virus_mse}. The relative square error is defined by: $||\bm{x-\hat{x}}||^2/||\bm{x}||^2$, where $\bm{x}$ is the ground truth image, and $\bm{\hat{x}}$ is the reconstructed image. The experiment shows that the VR-RK algorithm has a faster convergence rate and a better reconstruction accuracy compared with the randomized Kaczmarz algorithm and the iterative projection algorithms.   

\begin{table}[h!]
\begin{center}
\begin{tabular}{ |c|c|c|c|c|c|} 
 \hline
 ~ &VR-RK & RK & RAAR  & HIO \\ 
 \hline
 Error & 1.7540e-12 & 6.8635e-12 & 0.0307 & 0.1313 \\
 \hline
\end{tabular}
\end{center}
\caption{Reconstruction error comparison}
\label{table:mimi_virus_mse}
\vspace{-3mm}
\end{table}

\subsection{\textbf{Recovery efficiency under constraints}}
 \label{sec:robust}

\begin{figure}[ht]
    \centering
    \subfigure[Under $L_1$ constraint]{\includegraphics[width=0.48\textwidth, height=48mm]{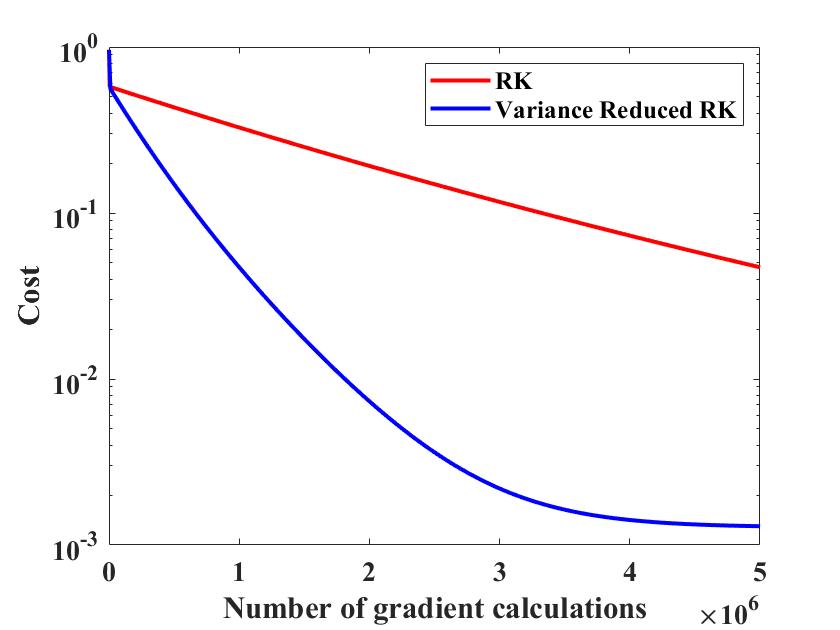}} 
    \subfigure[Under $L_2$ constraint]{\includegraphics[width=0.48\textwidth, height=48mm]{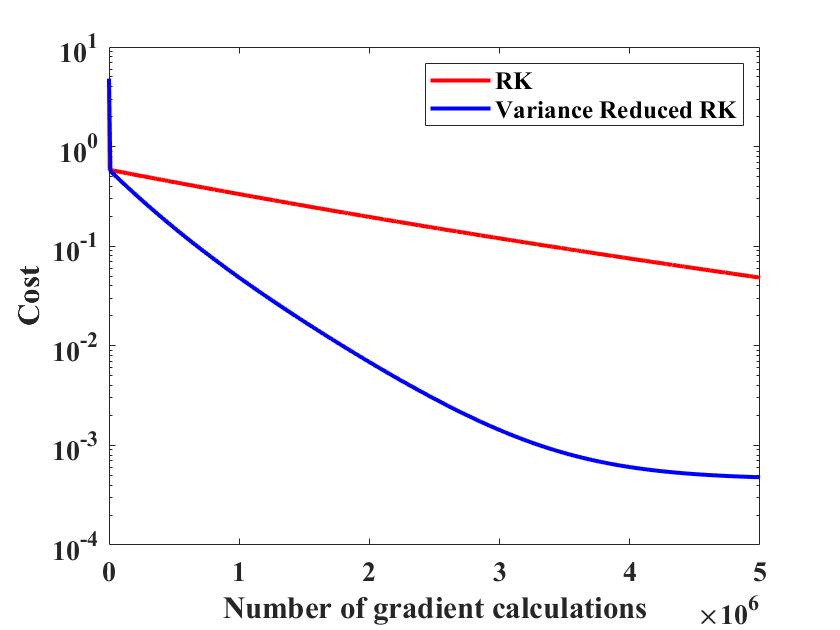}} 
\caption{Comparison of convergence rate}
    \label{fig:converg_comp}
\end{figure}

To further illustrate the convergence rate, we compare the VR-RK algorithm and the randomized Kaczmarz algorithm under $L_1$ and $L_2$ constraints on reconstructing the virus sample data. The cost function changes per iteration is shown in Figure~\ref{fig:converg_comp}. From the figure, the loss function decays faster in VR-RK than randomized Kaczmarz method. 

Considering that the single particle X-ray imaging data are influenced by the Poisson noise, and we examine the reconstruction accuracy at various noise levels, with $\epsilon$ from 0.005 to 0.1, and the measurement under the noise: $\bm{y=|Ax|}^2(1+\epsilon)$. 
 
Table~\ref{table:mimi_virus_noise_mse} shows the relative square error of reconstruction using different phase retrieval algorithms in various noise level. From Table~\ref{table:mimi_virus_noise_mse}, we can see that the VR-RK method outperforms other algorithms under noise. 

\begin{table}[h!]
\begin{center}
\begin{tabular}{ |c|c|c|c|c|c|} 
 \hline
 ~ &VR-RK-$L_2$ & VR-RK-$L_1$ & RAAR  & HIO \\ 
 \hline
 $\epsilon=0.1$ & 0.3687 & 0.3685 & 1.2502  & 0.7315 \\
 $\epsilon=0.05$ & 0.2438 & 0.2432 & 0.8775 & 0.5398 \\
 $\epsilon=0.01$ & 0.1007 & 0.1013 & 0.3860 & 0.3150 \\
 $\epsilon=0.005$ & 0.0707 & 0.0712 & 0.2703 & 0.2130 \\
 \hline
\end{tabular}
\end{center}
\caption{Relative square error comparison}
\label{table:mimi_virus_noise_mse}
\vspace{-3mm}
\end{table}

\subsection{\textbf{Results of the PR772 dataset}} \label{sec:pr772}

\begin{figure}[ht]
\setlength\abovecaptionskip{-0.1\baselineskip} 
    \centering
    \subfigure[]{\includegraphics[width=0.217\textwidth, height=30mm]{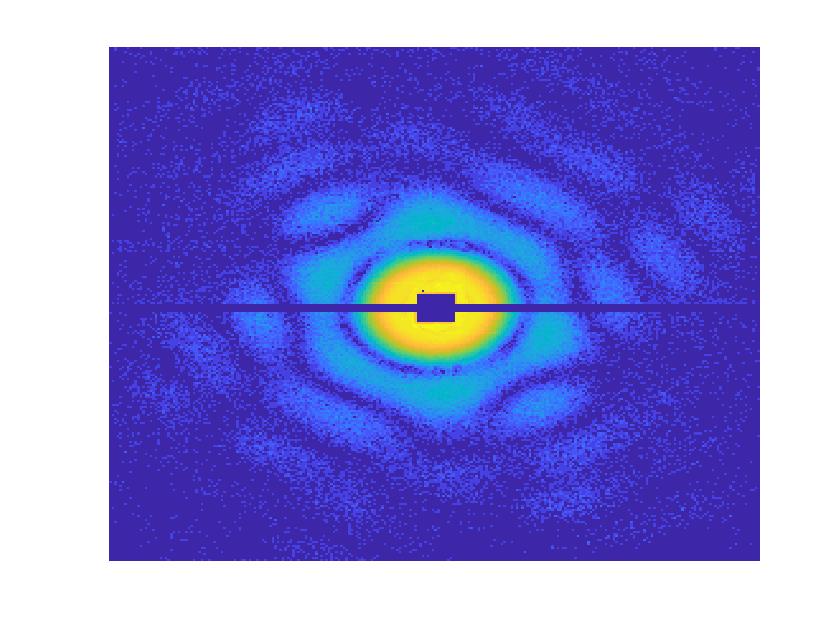}\hspace{0.1pt}} 
    \subfigure[]{\includegraphics[width=0.217\textwidth, height=30mm]{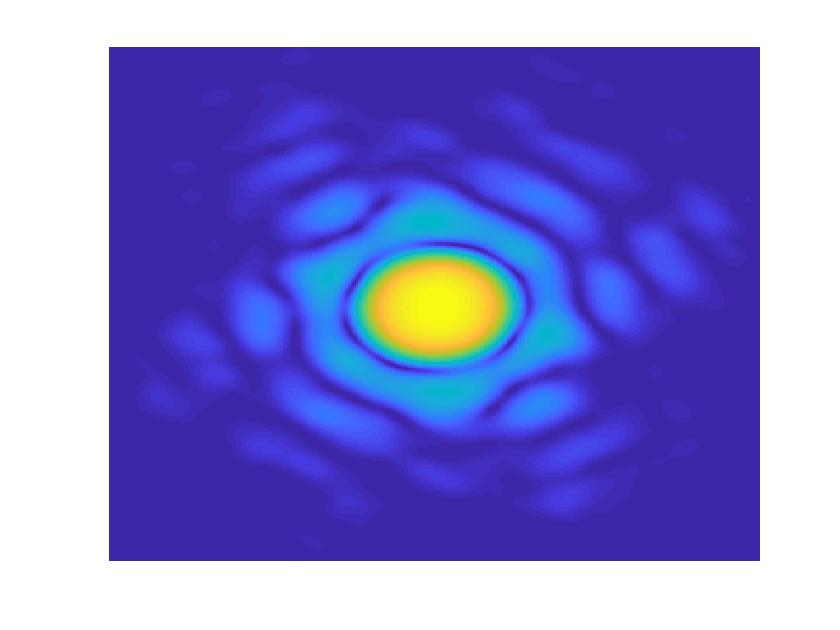}\hspace{0.1pt}}  
     \subfigure[]{\includegraphics[width=0.217\textwidth, height=30mm]{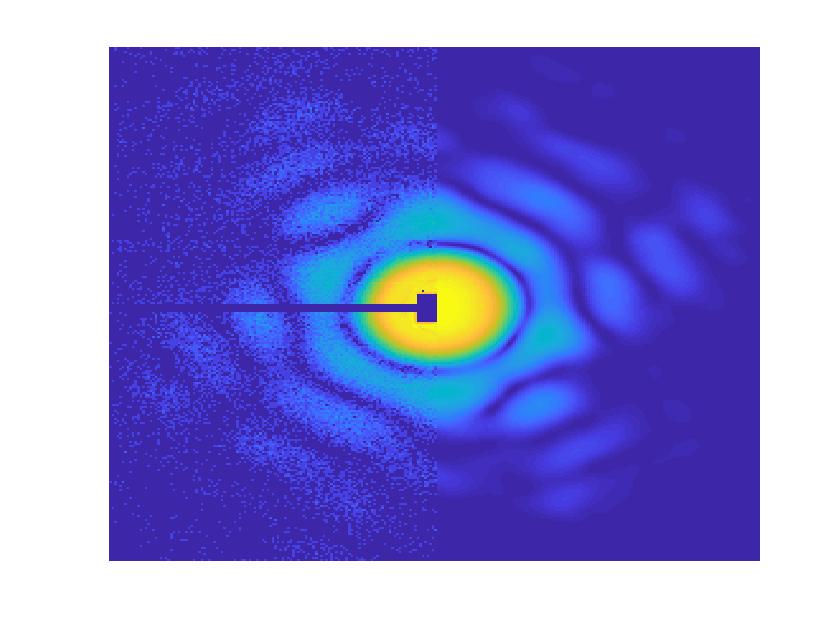}\hspace{0.1pt}}
    \subfigure[]{\includegraphics[width=0.217\textwidth, height=30mm]{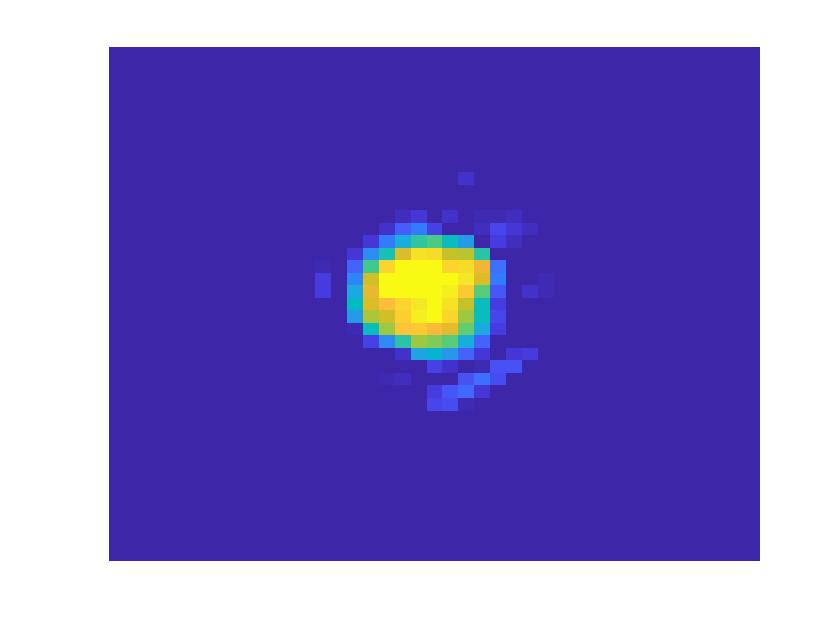}}  
\\
\vspace{-0.1in}
    \subfigure[]{\includegraphics[width=0.217\textwidth, height=30mm]{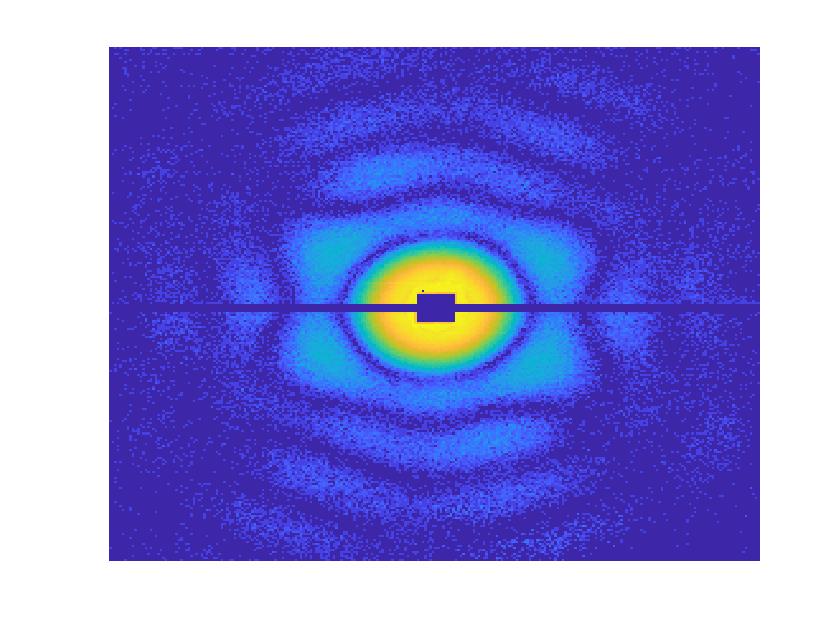}}  
    \subfigure[]{\includegraphics[width=0.217\textwidth, height=30mm]{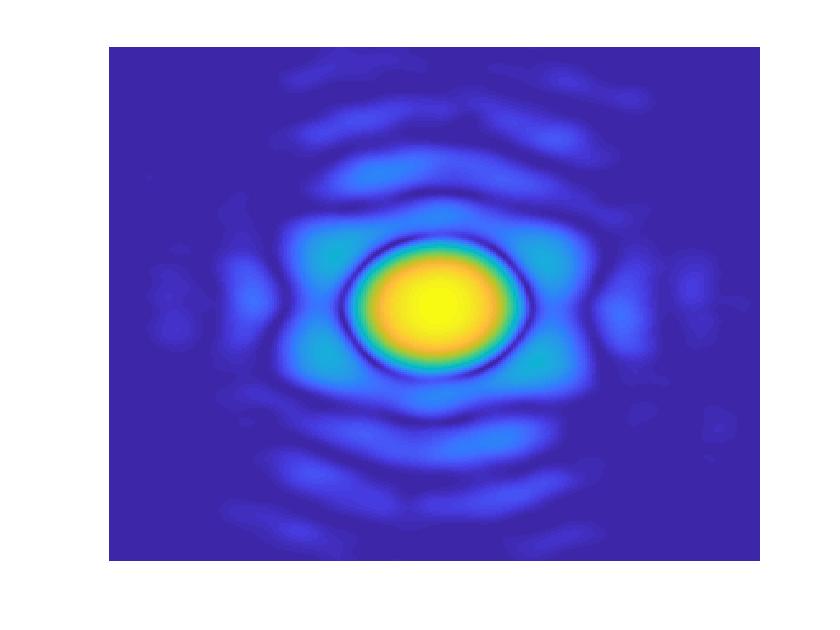}} 
    \subfigure[]{\includegraphics[width=0.217\textwidth, height=30mm]{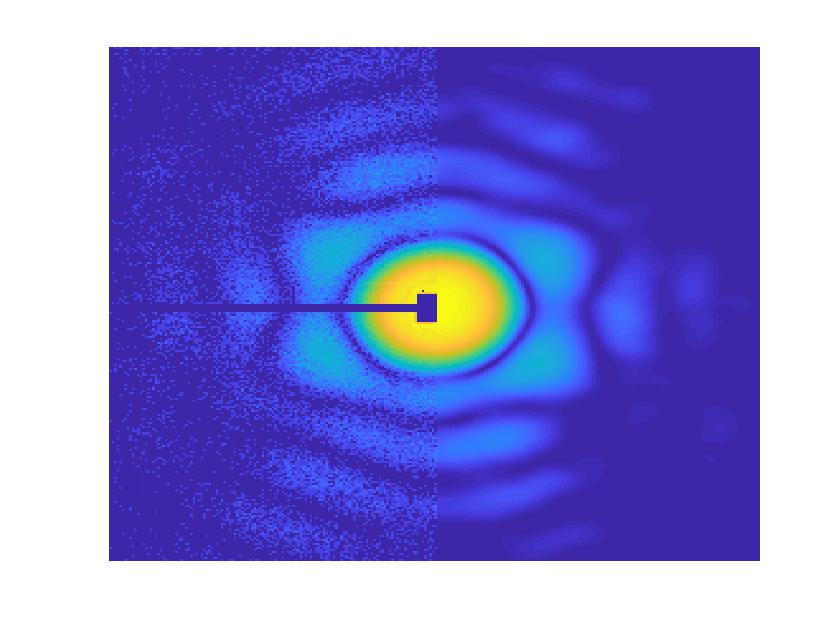}} 
    \subfigure[]{\includegraphics[width=0.217\textwidth, height=30mm]{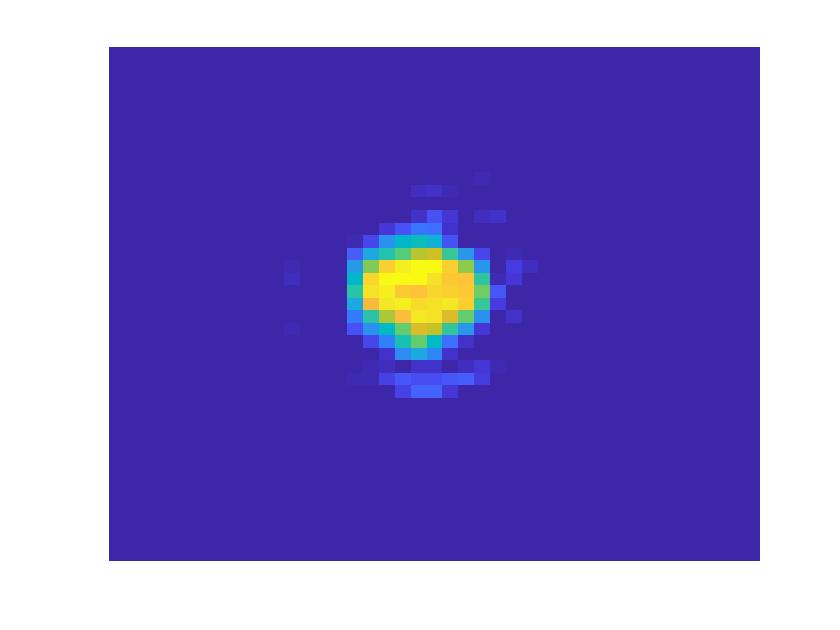}} 
\caption{PR772 single particle scattering pattern phase retrieval. (a) and (e) are two single particle diffraction pattern; (b) and (f) are the recovered diffraction pattern of (a) and (e) respectively; (c) and (g) show the comparison of the original and the recovered diffraction pattern, the left half is the original and the right half is the recovered; (d) and (h) are the real-space images reconstructed using VR-RK algorithm from (a) and (e).}
    \label{fig:pr772_re}
\end{figure}

We test the VR-RK algorithm on the PR772 particle dataset~(\cite{reddy2017}). The image size is $256 \times 256$ pixels, and the pixel values are scaled to the range of  [0, 255]. Illustration of the diffraction pattern of the single particle data is shown in Figure~\ref{fig:pr772_re}(a) and Figure~\ref{fig:pr772_re}(e).

For this dataset, the shrinkwrap method is applied to obtain a tight object support~(\cite{shi2019, marchesini2003}), and the square root of the diffraction intensities is used as a reference for the missing pixels during numerical iteration. An recovery example is shown in Figure~\ref{fig:pr772_re}, and more recovery examples are presented in the supplementary materials.

We use the VR-RK algorithm, RAAR and HIO methods to recovery the data, and classify the single-particle scattering pattern data and the non-single-particle scattering pattern data. We use the VR-RK for computation. There are 497 samples with labels in the validation set~(\cite{shi2019}). Among them, 208 are single-particle samples, and 289 are non-single-particle samples. We use ISOMAP for data compression and  clustering, and KNN for classification. We use 4-fold cross validation. The VR-RK has the best result. The AUCs of the binary classification results are listed as follows.

\begin{table}[h!]
\begin{center}
\begin{tabular}{ |c|c|c|c|c|} 
 \hline
 ~ &VR-RK  & RAAR  & HIO   \\ 
 \hline
AUC & 0.9501 & 0.9069 & 0.9231 \\
 \hline
\end{tabular}
\end{center}
\caption{AUC of binary classification}
\label{table:mimi_virus_auc}
\end{table}

From the results, we can see that the VR-RK method can help recover the data, and improve classification rate.

\section{\textbf{Conclusion}} \label{sec:conclusion}

In this chapter, we present the Variance-Reduced Randomized Kaczmarz (VR-RK) method for XFEL single particle phase retrieval. The VR-RK method is inspired by the randomized Kaczmarz method and the SVRG method. It is proposed in order to accelerate the convergence speed of the algorithm. Numerical results show that the VR-RK method has faster convergence rate and better accuracy under noises. Experiments on PR772 single particle X-ray imaging data show that the VR-RK method can help recover and classify particles.




\newpage 

\section{\textbf{Appendix}}

\begin{figure}[ht]
    \centering
    \subfigure[]{\includegraphics[width=0.49\textwidth, height=56mm]{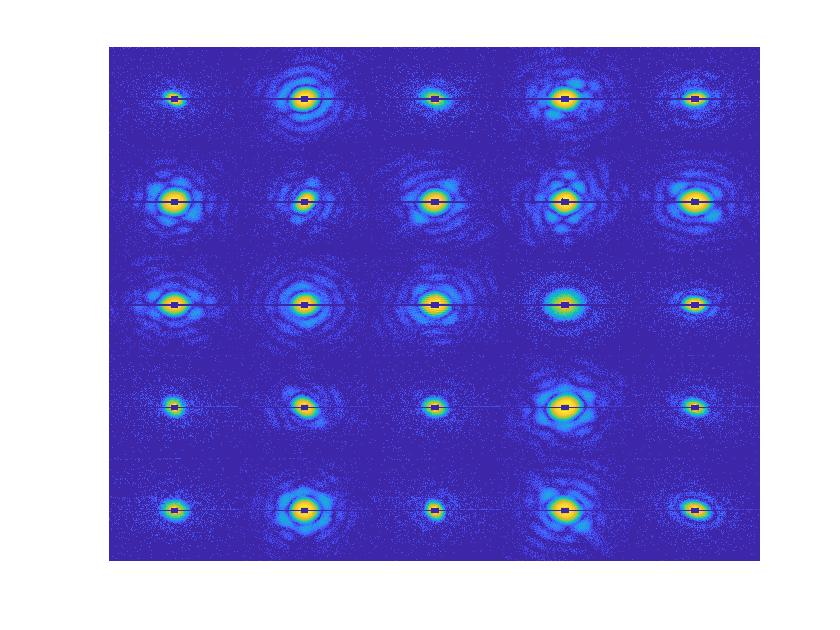}} 
    \subfigure[]{\includegraphics[width=0.49\textwidth, height=56mm]{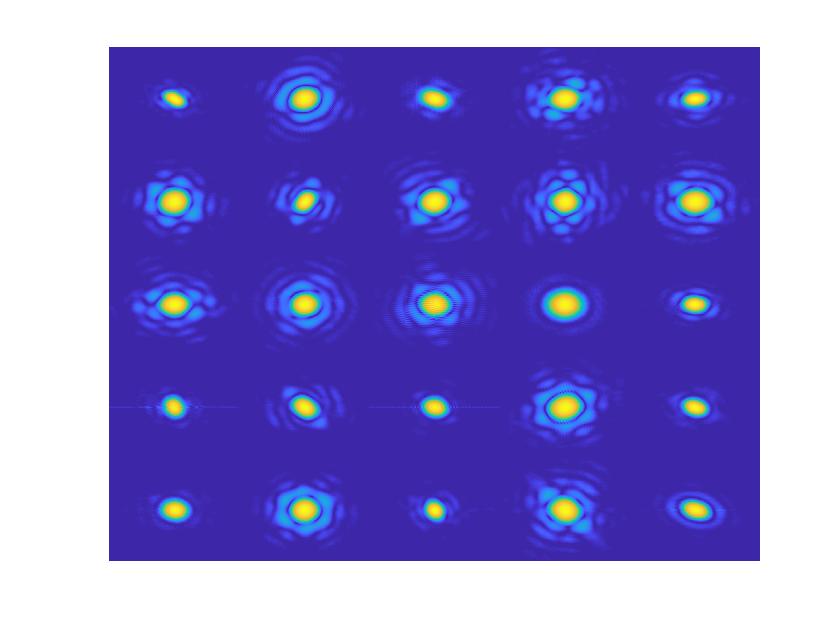}} 
      \subfigure[]{\includegraphics[width=0.49\textwidth, height=56mm]{output_img_5_5.jpg}} 
\vspace{-0.1in}
\caption{Phase retrieval of the PR772 dataset. (a) Original data diffraction pattern illustrations. (b) Recovered image diffraction pattern illustrations. (c) Recovered real space data illustrations. } 
    \label{fig:pr772_5x5}
\vspace{-3mm}
\end{figure}

For the PR772 dataset, further examples of phase retrieval recovery are shown here. Figure~\ref{fig:pr772_5x5}(a) and  Figure~\ref{fig:pr772_5x5}(b) are examples of 25 diffraction pattern samples reconstruction. Figure~\ref{fig:pr772_5x5}(c) shows the corresponding real space recovered images. 
\vspace{-0.15in}



\begin{figure}[h]
    \centering
    \includegraphics[width=0.73\textwidth, height=87.2mm]{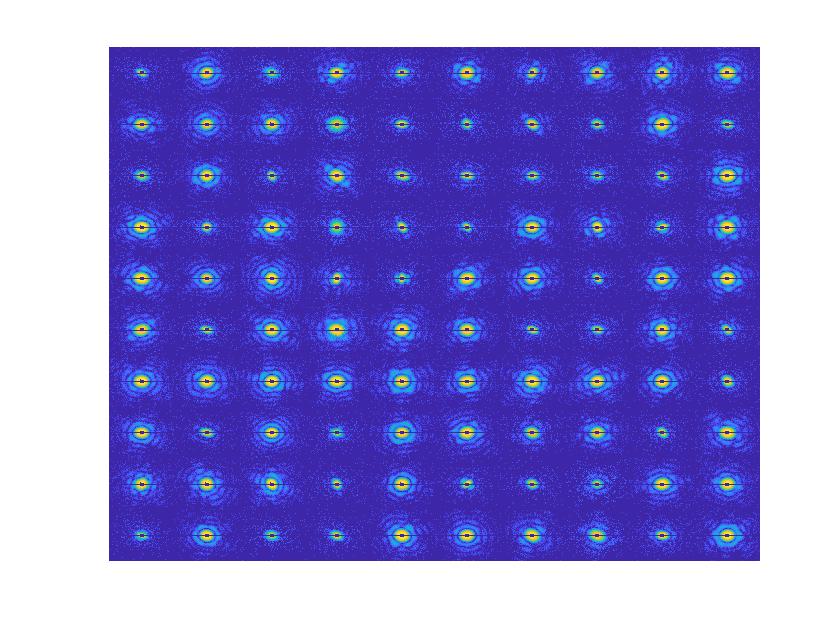}
    \vspace{-0.3in}
\caption{Original data diffraction pattern illustrations}
    \label{fig:diff_100_pr772}
\end{figure}
\begin{figure}[h]
    \centering
    \includegraphics[width=0.73\textwidth, height=87.2mm]{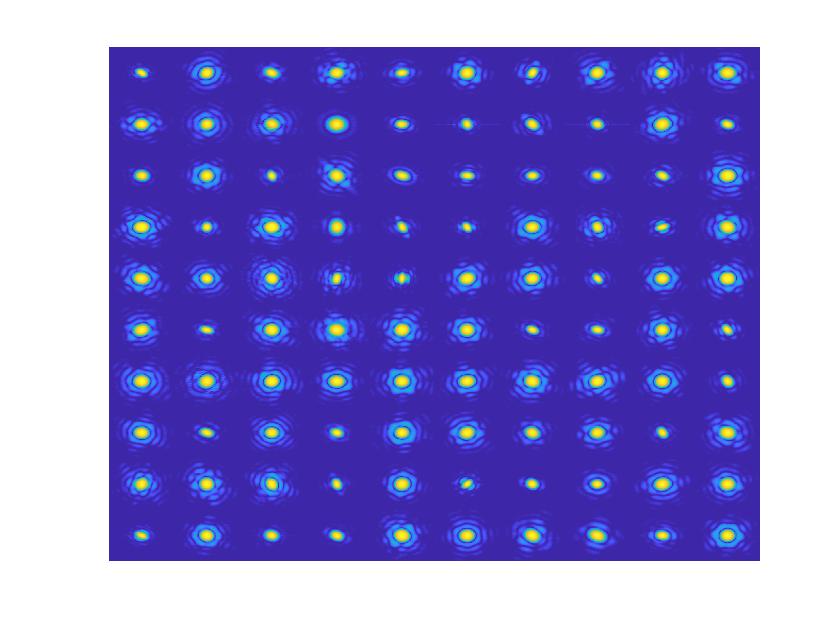}
    \vspace{-0.3in}
\caption{Recovered image diffraction pattern illustrations}
    \label{fig:recov_100_pr772}
\end{figure}
\begin{figure}[h]
    \centering
    \includegraphics[width=0.73\textwidth, height=87.2mm]{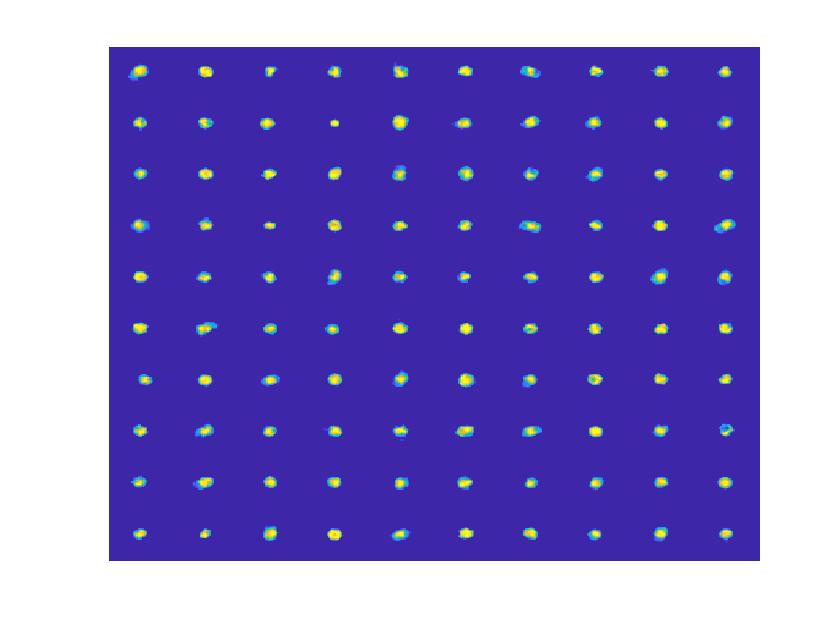}
\vspace{-0.3in}
\caption{Recovered real space data illustrations}
    \label{fig:recov_real_100_pr772}
\end{figure}

\newpage
Figure~\ref{fig:diff_100_pr772} and  Figure~\ref{fig:recov_100_pr772} are examples of 100 diffraction pattern samples reconstruction. Figure~\ref{fig:recov_real_100_pr772} shows the corresponding real space recovered images. 

\vspace{-0.14in}

\vspace{-0.25in}

\vspace{-0.25in}

\vspace{-0.2in}

\newpage




%
%




\newpage
\addcontentsline{toc}{section}{\textbf{Reference}}

\bibliographystyle{plainnat}
\bibliography{sample.bib}

\end{document}